\documentclass[%
 reprint,
 amsmath,amssymb,
 aps,
pra,
]{revtex4-2}

\usepackage{graphicx}% Include figure files
\usepackage{dcolumn}% Align table columns on decimal point
\usepackage{bm}% bold math
\usepackage{hyperref}% add hypertext capabilities
\begin{document}

\preprint{APS/PRA}

\title{Temperature shift suppression scheme\\for two-photon two-color rubidium vapor clocks}% Force line breaks with \\

\author{Tin Nghia Nguyen}
 \affiliation{Department of Physics, University of Colorado, Boulder, Colorado 80309-0390, USA}%Lines break automatically or can be forced with \\
\author{Thomas R. Schibli}%
 \email{trs@colorado.edu}
\affiliation{Department of Physics, University of Colorado, Boulder, Colorado 80309-0390, USA}%

\date{\today}% It is always \today, today,

\begin{abstract}
We propose a new scheme for interrogating a warm rubidium vapor using two different clock lasers. Performance-wise, this approach is distinctly different from the recently proposed two-color two-photon rubidium clocks as our scheme does not trade off the AC Stark suppression against an increased sensitivity to the cell-temperature/pressure. Instead, our approach compensates all, the AC-Stark shift and the temperature \& pressure-induced frequency shifts. The proposed scheme also makes use of the modulation transfer technique, which enables a two-orders of magnitude increase in the signal-to-noise ratio compared to traditional clocks that rely on fluorescence measurements. 
\end{abstract}

\maketitle

%\tableofcontents

\section{\label{sec:Intro}Introduction}

Ultra-stable optical atomic clocks are crucial for applications involving positioning, navigation and timing (PNT), and have found their place at the heart of many important scientific experiments. To date, the most stable atomic clocks reach fractional frequency instabilities of the order of 10$^{-19}$ and are typically based on single (or few) ions or on a large number of neutral atoms trapped in an optical lattice. Such clocks typically achieve their stability by probing doubly-forbidden optical transitions of very narrow line-widths~\cite{Marti2018,Brewer2019}. Both types of clocks are housed inside elaborate vacuum systems and require laser cooling of single ions or atomic ensembles down to near absolute zero temperatures. Because of this, such clocks typically occupy cubic meters of space and require a highly skilled group of people to maintain them.

Optical atomic clocks based on warm atomic (or molecular) vapors, on the other hand, neither require a vacuum system nor laser cooling, and their size is typically best measured in liters, rather than cubic meters. These clocks employ a vapor cell that is either slightly heated or cooled beyond room temperature to maintain a desirable vapor pressure. However, such clocks only deliver fractional instabilities at the order of 10$^{-12}$ to 10$^{-15}$~\cite{Phelps2018, Hou2014, Maurice2020, Terra2016, Martin2018, Hilico1998}. This ‘poor’ stability is not only due to the much broader linewidths of the clock transition (e.g., a few 100~kHz for the 5S$_{1/2}$~$\rightarrow$~5D$_{5/2}$ rubidium two photon transition), but also due to AC-Stark and pressure-induced shifts. These shifts are mainly driven by intensity fluctuations of the probe-laser and temperature variations of the vapor cell. Reaching fractional frequency instabilities beyond 10$^{-15}$ impose technologically unrealistic long-term constraints on the laser power and the cell temperature. 

A new scheme was recently proposed to significantly reduce AC-Stark-induced shifts to the 5S$_{1/2}$~$\rightarrow$~5D$_{5/2}$ two photon transition in Rubidium vapor clocks by using two counter-propagating laser beams with one laser at a few GHz blue-detuned from the 5S$_{1/2}$~$\rightarrow$~5P$_{3/2}$ transition at 780~nm and the other laser at 776~nm and maintaining a suitable ratio of the laser powers~\cite{Gerginov2018}. This scheme also includes a two-orders of magnitude increase in the signal to noise ratio (SNR) that could allow a higher fractional stability at short time scales. Unfortunately, as discussed later in this letter, this two-color scheme also adds a first-order contribution from the optical Doppler shift to the clock transition due to unequal laser frequencies, which makes this scheme generally more sensitive to the cell temperature variations. This could explain why the short-term stability of this method was found to be no different from that of common rubidium two-photon clocks. Here we propose for the first time to employ the two-color method to cancel the residual Doppler shift against frequency shifts induced by cell temperature variations. Due to the new flexibility offered by this two-color method, we can achieve a full cancellation of the first-order residual Doppler shift and the temperature-induced pressure shifts without requiring special gas mixtures. This could, for the first time, lead to a vapor clock that is insensitive to both, optical power fluctuations and temperature/pressure fluctuations of the atomic vapor. Overall, this would enable liter-sized optical clocks with a stability of the previous-generation cold atomic clocks. 

\section{\label{sec:RbC}The two-photon Rubidium vapor clock}

\subsection{\label{sec:onecolor}One-color scheme}
\begin{figure}[b]
	\includegraphics[width = 0.4\textwidth]{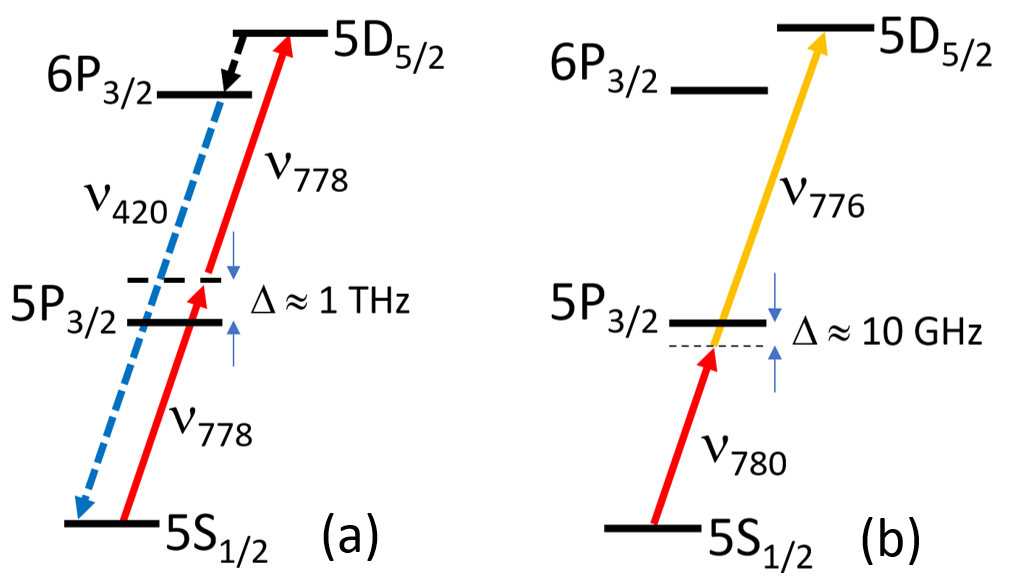}
	\caption{\label{fig:schemes} Energy diagram of the two-photon Rb clock using the 5S$_{1/2}$~$\rightarrow$~5D$_{5/2}$ transition following (a) the common single-color scheme with single wavelength of 778.1~nm and (b) the two-color scheme with light at 776-nm and 780-nm wavelengths for AC-Stark shift cancellation and temperature shift suppression upgrades.}
\end{figure}

The state-of-the-art Rubidium vapor clocks employ the two-photon 5S$_{1/2}$~$\rightarrow$~5D$_{5/2}$ transition as the transition of choice~\cite{Phelps2018, Terra2016, Martin2018, Hilico1998, Newman2021}. By absorbing two counter-propagating photons from a single laser at a wavelength of 778.1~nm, the Rb atoms can be excited to the 5D$_{5/2}$ state, from where they then decay to the 6P$_{3/2}$ state and back to the ground state emitting a blue photon at wavelength of 420~nm as shown in Fig.~\ref{fig:schemes}a. By detecting these blue photons with a photomultiplier tube (PMT), the frequency of the 778-nm laser can be tuned to matched exactly half of the frequency of the 5S$_{1/2}$~$\rightarrow$~5D$_{5/2}$ transition. The most common source of 778.1~nm light for these clocks is from the frequency doubled light of a telecom Erbium fiber laser at 1556.2~nm. Due to the maturity of telecom components, these Erbium fiber lasers reliably achieve low intensity noise and ultra-low phase noise with laser linewidths narrower than 1~kHz. This ensure a sufficient SNR in such clocks to get the fractional stability at the level of 10$^{-13}$ at 1~s, and 10$^{-14}$ after long averaging times~\cite{Bigelow2018}. To improve beyond that, various steps need to be taken to increase the SNR and to minimize or eliminate the sources of shifts in frequency, namely the AC-Stark shift, temperature shifts, Zeeman shift, and Helium collisional shift. While the Zeeman shift and the Helium collisional shift can be mitigated by designing a properly demagnetized multi-layer magnetic shield and employing a cell made with Aluminosilicate glass, respectively, to achieve 10$^{-15}$ fractional stability, the other two shifts require strict control of the laser power and the temperature of the vapor cell down to the microwatt-level power fluctuation and sub-millikelvin temperature fluctuation over one-day averaging time~\cite{Martin2018}. These strict requirements make it impractical to maintain the clock stability over long periods of time and would drastically increase the size, weight, and power (SWaP) of such clocks. 

\subsection{\label{sec:twocolor}Two-color scheme}
Another clock scheme was proposed by Gerginov, Beloy and Perrella et. al that uses two counter-propagating laser beams at different colors to interrogate the atoms: one laser is a few GHz blue-detuned from the 5S$_{1/2}$~$\rightarrow$~5P$_{3/2}$ transition at 780~nm and the other laser is at 776~nm~\cite{Gerginov2018, Perrella2019}. This scheme features a many-fold increase in SNR due to the smaller detuning of one of the lasers from the 5S$_{1/2}$~$\rightarrow$~5P$_{3/2}$ transition (a few GHz compared to 1~THz of the 1-color method as shown in Fig.~\ref{fig:schemes}), its use of highly efficient Silicon photodiodes to detect the laser transmission instead of the PMT, and a larger interaction volume with the atoms. The scheme was also designed to suppress the AC-Stark shift by employing the difference in the sign of the polarizability between the 780~nm and 776~nm excitation pathways, and by maintaining a suitable ratio of the laser powers. It was found by Gerginov and Beloy that the AC-Stark shift can be completely annihilated if the following condition is satisfied

\begin{equation}
\frac{I_{780}}{I_{776}} = (0.0656)[1-(8.06\times 10^{-3})\Delta-(3.19\times 10^{-6})\Delta^2],
\label{eqn:Intensity_ratio}
\end{equation}
where $I_{780}$, and $I_{776}$ are the intensities of the 780-nm and 776-nm laser beam, and $\Delta$ (in $2\pi \times$~GHz) is the detuning of the 780-nm laser from the 5S$_{1/2}$~$\rightarrow$~5P$_{3/2}$ transition which can take on both positive and negative values, i.e. both blue and red detunings. Unfortunately, a drawback of using two different colors to interrogate the atoms is that the Doppler effect cannot be fully eliminated. This results in a Voigt transition line shape~\cite{Perrella2013} and leads to a residual Doppler broadening of the transition line to a couple of MHz at 90~$^\circ$C, and it leads to and a residual net-Doppler shift, which is given by

\begin{equation}
\nu_{780}+\nu_{776}-\nu_{fg} \approx \frac{1}{4\ln 2}\frac{(\nu_{780}-\nu_{776})\nu_{780}(\bar{v}/c)^2}{\nu_{780}-\nu_{ig}}
\label{eqn:Pulling_effect}
\end{equation}

where $\nu_{776}$ and $\nu_{780}$ are the frequencies of the 776~nm and 780~nm lasers, respectively, $\nu_{fg}$ is the natural transition frequency between the ground state 5S$_{1/2}$ and the excited state 5D$_{5/2}$, while $\nu_{ig}$ is the transition frequency between the ground state 5S$_{1/2}$ and the intermediate state 5P$_{3/2}$, $\bar{v} = \sqrt{8k_BT\ln 2/m}$ is proportional to the average speed of the atoms, with $k_B$ being the Boltzmann constant, $T$ being the absolute temperature of the atoms, and $m$ being the mass of a Rb atom for large detunings of the 780-nm laser from the intermediate level~\cite{Bjorkholm1976}. It is clear that the pulling effect is proportional to $\bar{v}^2$ and the temperature $T$. It should be noted that the sign of the shift depends on the difference between the frequencies of the 780-nm laser and the 5S$_{1/2}$~$\rightarrow$~5P$_{3/2}$ transition, and that the pulling effect is strongest for small detunings between these two. The latter two of the remarks can be seen clearly in the region of low detunings in Fig.~2 in the paper presented by Perrella et. al.~\cite{Perrella2013}. Moreover, by replacing the two wavelengths of 780~nm and 776~nm with a single wavelength at 778~nm in Eqn.~\ref{eqn:Pulling_effect}, as for the case of one-color two-photon Rubidium clocks, the pulling effect is fully cancelled out, as expected.

For the AC-Stark shift cancellation method proposed by Gerginov and Beloy, since the 780-nm laser is blue-detuned from the 5S$_{1/2}$~$\rightarrow$~5P$_{3/2}$ transition, the pulling effect decreases the transition frequency with increasing temperature similar to the Rb-Rb collisional shift and the black-body radiation shift. A quick calculation of the pulling effect gives a temperature coefficient of -170~Hz/$^\circ$C for a 10~GHz blue detuned 780-nm laser, and -860~Hz/$^\circ$C for a 2~GHz blue detuned laser at 100~$^\circ$C cell temperature, respectively. These are at the same order of magnitude as the temperature coefficient of the Rb-Rb collisional shift of -925~Hz/$^\circ$C to -423~Hz/$^\circ$C at 100~$^\circ$C~\cite{Phelps2018, Newman2021}. Thus, the proposed 2-color scheme with AC-Stark shift cancellation increases the clock’s temperature coefficient and worsens the clock stability for a given temperature fluctuation. In other words, the scheme proposed by Gerginov and Beloy scheme exchanges trades the AC-Stark shift for a higher temperature shift and, which results in unchanged overall stability.

For the clock scheme by Perrella et. al, the 780-nm laser is red-detuned by 1.5~GHz from the intermediate state, which could help cancelling the Rb-Rb collisional shift. However, since the detuning is relatively small and the pulling effect is inversely proportional to this detuning, the pulling effect heavily outweighs the Rb collisional shift at low temperatures. Simple calculations give the a pulling effect shift frequency of 430~kHz at 90~$^\circ$C, which is much larger than the Rb-Rb collisional shift of -1.82~kHz. The pulling effect temperature coefficient can be computed to be 1.2~kHz/$^\circ$C, which is slightly larger than the temperature coefficient of Rb-Rb collisional shift mentioned earlier at the relatively the same  temperature. At higher temperatures, the Rb-Rb collisional shift temperature coefficient increases at a faster pace than the pulling effect coefficient due to the exponential increase in Rb vapor pressure. This means that at some temperature higher than 90~$^\circ$C, the coefficients of these shifts would be the same but opposite,  creating a local maximum in the frequency shift plotted against the cell temperature curve. This local maximum would allow a lower temperature-induced frequency shift, and therefore, simpler temperature-stabilizing stabilization methods for the two-color clocks. Hence, we came up with an improved better version of the two-color clock scheme to take advantage of this idea. 

\subsection{\label{sec:newscheme}The new scheme}
\begin{figure*}
	\includegraphics[width = \textwidth]{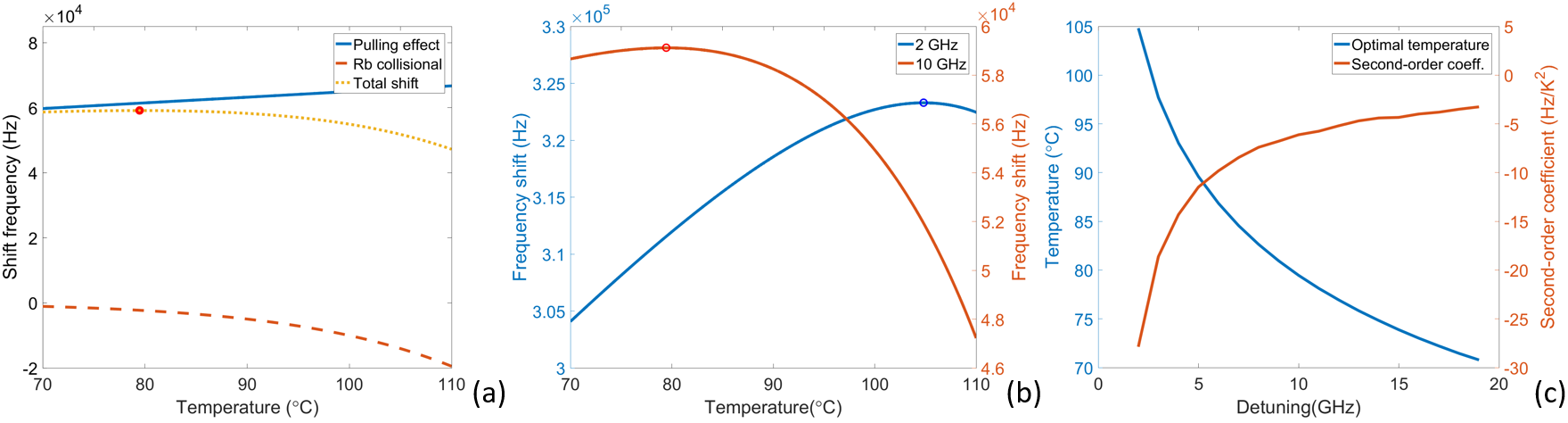}
	\caption{\label{fig:new_scheme} (a) Contributions of the pulling effect and the collisional shift as a function of the vapor temperature when the 780-nm laser is detuned by 10~GHz below the 5S$_{1/2}$~$\rightarrow$~5P$_{3/2}$ transition. The red circle indicates the location of a local maximum in the total temperature shift. (b) Net-frequency shift as a function of temperature when the 780-nm laser is detuned 2~GHz (blue) and 10~GHz (amber) below the 5S$_{1/2}$~$\rightarrow$~5P$_{3/2}$ transition. The blue and red circles indicate the locations of the local maximum of each curve, where the temperature coefficient vanishes. (c) The optimum temperature and the second-order shift coefficient at the optimum temperature as a function of the detuning of the 780-nm laser.}
\end{figure*}

We propose another scheme for the two-photon vapor Rubidium clock similar to the methods proposed by Gerginov and Beloy and by Perrella et. al., but with a twist. In our proposed method, we make the 780-nm laser red-detuned by 10~GHz from the 5S$_{1/2}$~$\rightarrow$~5P$_{3/2}$ transition. This makes the pulling effect take on the sign opposite to the Rb-Rb collisional shift and the black-body radiation shift, making it a perfect candidate to counter those temperature-induced shifts. The higher detuning compared to the scheme by Perrella et. al allows the clock to be operated at a lower optimal temperature for reduced temperature-induced frequency shifts. As shown in Fig.~\ref{fig:new_scheme}a, the frequency pulling effect can be used to suppress the Rb collisional shift, the main contributor to the total temperature shift, due to their difference in signs. As a result, we find a local maximum in the temperature-induced shifts. The Rb-Rb collisional shift was calculated from the pressure vs. temperature curve of Rubidium, while accounting for the approximate factor of two between the measured vapor pressure shift of -27~kHz/mTorr for the 5S$_{1/2}$~$\rightarrow$~5D$_{5/2}$ of $^{85}$Rb isotope in a vapor cell with natural Rb compared to the collisional shift in highly-enriched $^{87}$Rb~\cite{Martin2018, Zameroski2014}. By operating the vapor cell at a temperature around this local maximum, the dependency of the clock frequency on temperature can be described as $\Delta\nu=a(T-T_{opt})^2$, where `$a$' is the second-order temperature coefficient plotted in Fig.~\ref{fig:new_scheme}c for various detunings of the 780 nm laser. For 10 GHz detuning of the 780 nm laser, `$a$' takes on a value of -6.12~Hz/K$^2$, meaning that a temperature drift of as much as 100~mK around the optimal temperature of 79.5~$^\circ$C would only result in a net-frequency change of 0.061~Hz or a contribution to the fractional instability at $8\times10^{-17}$. This is indeed orders of magnitude smaller compared to what can be achieved by today's one-color clocks' sophisticated temperature stabilizing schemes. A drawback of this scheme is that the optimal temperature depends on the detuning of the 780-nm laser as shown in Fig.~\ref{fig:new_scheme}b. Therefore, the 780-nm laser frequency needs to be stabilized relative to another laser probing the 5S$_{1/2}$~$\rightarrow$~5P$_{3/2}$ transition, which can be addressed by either sending a second probing laser through the same cell locking it to the 5S$_{1/2}$~$\rightarrow$~5P$_{3/2}$ transition, and locking the beat note between this laser and the 780-nm laser to 10~GHz, or by generating a 10~GHz offset single side band from the 780-nm probe laser, and locking the side band to the said transition.

\begin{figure}[b]
	\includegraphics[width = 0.4\textwidth]{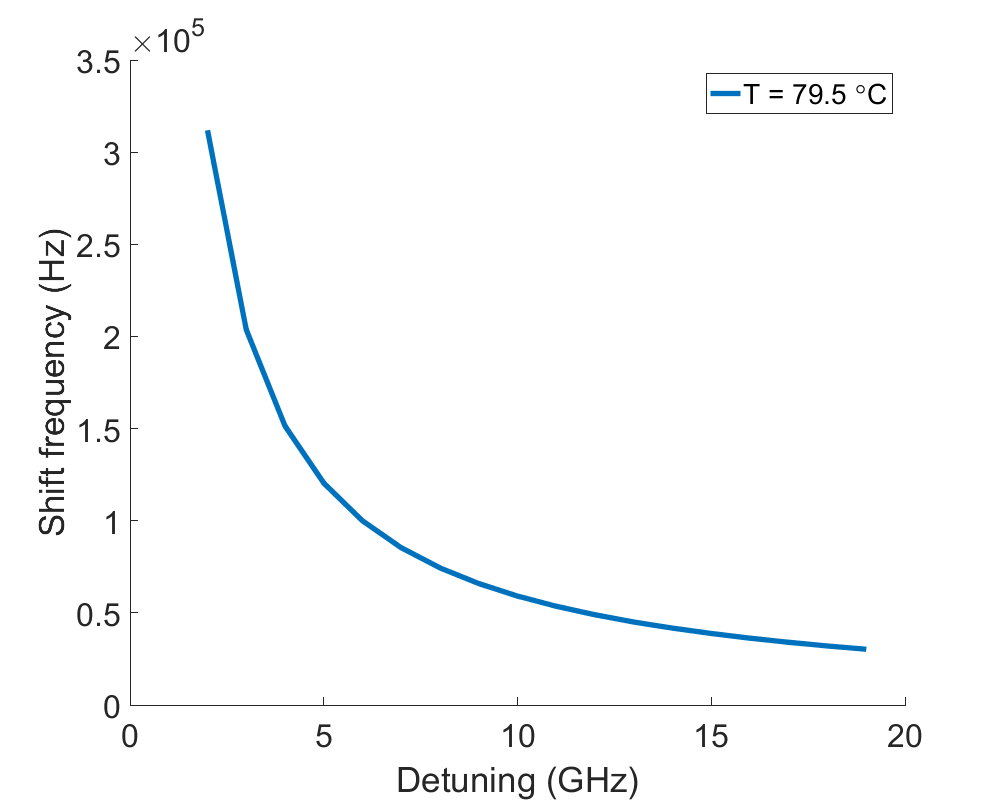}
	\caption{\label{fig:shift_detuning} The shift in transition line as a function of detunings of the 780-nm laser from the 5S$_{1/2}$~$\rightarrow$~5P$_{3/2}$ transition at the optimal temperature for a 10~GHz detuning.}
\end{figure}

\section{\label{sec:Discussion}Discussion}
The temperature shift suppression scheme closely resembles the clock scheme of Gerginov et. al., which not only allows for a many-fold increase in SNR compared to one-color Rubidium vapor clocks, but also for the AC-Stark shift suppression scheme to be incorporated to result in a scheme that suppresses both, the AC-Stark shift and the pressure-induced shifts. However, the power ratio of the laser beams, $I_{780}/I_{776}$, will have to be changed from 0.0602 for a 10~GHz blue-detuned 780-laser to 0.0709 for a 10~GHz red-detuned 780-nm laser following Eqn.~\ref{eqn:Intensity_ratio}. For an input power of the 780-nm laser of $P_{780} = 1.2$~mW, a power of $P_{776} = 16.93$~mW is needed to provide the required power ratio of $P_{780}/P_{776} = 0.0709$ for the AC-Stark shift cancellation to take effect. Testing with 1\% of power fluctuation in one of the laser diodes around the optimal power values for a beam radius of 1~mm, we found an induced frequency shift of -69.6~Hz or $1.8\times10^{-13}$ fractional instability. Therefore, to achieve a noise floor of, say $3\times10^{-15}$ at 100~s, the absolute power of each laser diode needs to be stabilized to a fractional level of $1.66\times10^{-4}$. This can be done indeed, as power instabilities at the level of $6\times10^{-7}$ at 100~s was achieved using an acousto-optic modulator for an external cavity diode laser at 7~mW of total output power~\cite{Tricot2018}.

An apparent trade-off of this scheme is that the detuning of the 780 nm laser from the 5S$_{1/2}$~$\rightarrow$~5P$_{3/2}$ transition needs to be locked relatively well to 10~GHz as a change in the detuning leads to a frequency shift to the transition line. However, this shift in frequency is drastically reduced with higher detunings of the 780-nm laser, as shown in Fig.~\ref{fig:shift_detuning}. For a red detuning of 10~GHz, the change in frequency shift is roughly -12.33~kHz per 1~GHz change of the frequency of the 780-nm laser. This means that to attain a fractional instability of 10$^{-15}$, the 780-nm laser detuning needs to be stabilized down to 60~kHz level, which can be done by phase locking the beat note between the 780-nm laser and a laser probing the 5S$_{1/2}$~$\rightarrow$~5P$_{3/2}$ transition to a low noise, low drift oscillator at 10~GHz.

Last but not least, it is important to point out that the arguments and simulations above assumed that the temperature of the atomic vapor and the temperature controlling the vapor pressure to be the same. In reality, the vapor pressure is dictated by the temperature of the coldest spot on the cell wall, usually at the stem of the glass cell, while the temperature of the atomic vapor is given by the cell body temperature. Unfortunately, these temperatures are not the same as observed in a vapor glass cell with long stems~\cite{Calosso2012}. However, these temperatures could be made to be highly or monotonously correlated by having better insulation between the entire glass cell and the environment, and by controlling both temperatures with only one active temperature stabilization loop~\cite{Calosso2012,Micalizio2012}. Then, the temperature setpoint can be varied to map out the parabolic region of the temperature-induced frequency shift. The existence of the parabolic region is always guaranteed because the temperature coefficient of the frequency pulling effect mostly stays the same while that of the Rb-Rb collisional shift increases with temperature due to the increase in vapor pressure as shown in Fig.~\ref{fig:new_scheme}a. As a result, the clock can be made to be insensitive to temperature fluctuations by adjusting the setpoint temperature to the vertex of the parabolic temperature curve, which might be slightly different from the optimal temperature calculated above.

\section{\label{sec:Conclusion}Conclusion}
In this letter, we present an analysis of temperature-induced shifts of the proposed two-color clock schemes and explain that why, despite the higher SNR presented by the scheme, the short-term stability of this method was found to be no different from that of common rubidium two-photon clocks. It was found that these previous schemes had a significantly higher temperature coefficient than that of one-color rubidium vapor clock schemes, mainly due to the residual first-order Doppler shift. Here, we propose another clock scheme to employ this residual Doppler shift to suppress the Rb-Rb collisional shifts at the expense of introducing another laser to probe the 5S$_{1/2}$~$\rightarrow$~5P$_{3/2}$ transition. By phase-locking the 780-nm laser to an auxiliary laser probing the 5S$_{1/2}$~$\rightarrow$~5P$_{3/2}$ transition to better than 60 kHz uncertainty, the temperature-induced shift can be made to be below 10$^{-15}$ fractional instability level. This feature, together with the inherently higher SNR and the AC-Stark shift suppression scheme would enable vapor clocks to attain the frequency stability of the previous-generation table-top cold atomic clocks.
\bibliography{apssamp}% Produces the bibliography via BibTeX.

\end{document}